\documentstyle[twoside,epsf]{article}

\catcode`\@=11
\long\def\@makefntext#1{
\protect\noindent \hbox to 3.2pt {\hskip-.9pt  
$^{{\eightrm\@thefnmark}}$\hfil}#1\hfill}		%CAN BE USED 

\def\@makefnmark{\hbox to 0pt{$^{\@thefnmark}$\hss}}	%ORIGINAL 
	
\def\ps@myheadings{\let\@mkboth\@gobbletwo
\def\@oddhead{\hbox{}
\rightmark\hfil\eightrm\thepage}   
\def\@oddfoot{}\def\@evenhead{\eightrm\thepage\hfil
\leftmark\hbox{}}\def\@evenfoot{}
\def\sectionmark##1{}\def\subsectionmark##1{}}

\oddsidemargin=\evensidemargin
\newcounter{sectionc}\newcounter{subsectionc}\newcounter{subsubsectionc}
\renewcommand{\section}[1] {\vspace{12pt}\addtocounter{sectionc}{1} 
\setcounter{subsectionc}{0}\setcounter{subsubsectionc}{0}\noindent 
	{\tenbf\thesectionc. #1}\par\vspace{5pt}}
\renewcommand{\subsection}[1] {\vspace{12pt}\addtocounter{subsectionc}{1} 
	\setcounter{subsubsectionc}{0}\noindent 
	{\bf\thesectionc.\thesubsectionc. {\kern1pt \bfit #1}}\par\vspace{5pt}}
\renewcommand{\subsubsection}[1] {\vspace{12pt}\addtocounter{subsubsectionc}{1}
	\noindent{\tenrm\thesectionc.\thesubsectionc.\thesubsubsectionc.
	{\kern1pt \tenit #1}}\par\vspace{5pt}}

\newcommand{\textlineskip}{\baselineskip=13pt}
\newcommand{\smalllineskip}{\baselineskip=10pt}

\def\eightcirc{
\begin{picture}(0,0)
\put(4.4,1.8){\circle{6.5}}
\end{picture}}
\def\eightcopyright{\eightcirc\kern2.7pt\hbox{\eightrm c}} 

\newcommand{\copyrightheading}[1]
	{\vspace*{-2.5cm}\smalllineskip{\flushleft
        {\footnotesize Physics World vol. 12 (October 1999) 21-22 #1}\\
       {\footnotesize Los Alamos electronic archives: physics/99100nn #1}\\
        {\footnotesize $\eightcopyright$\,
        1999 H.C. Rosu
        }\\
	 }}

\newcounter{itemlistc}
\newcounter{romanlistc}
\newcounter{alphlistc}
\newcounter{arabiclistc}

\def\@citex[#1]#2{\if@filesw\immediate\write\@auxout
	{\string\citation{#2}}\fi
\def\@citea{}\@cite{\@for\@citeb:=#2\do
	{\@citea\def\@citea{,}\@ifundefined
	{b@\@citeb}{{\bf ?}\@warning
	{Citation `\@citeb' on page \thepage \space undefined}}
	{\csname b@\@citeb\endcsname}}}{#1}}

\newif\if@cghi
\def\cite{\@cghitrue\@ifnextchar [{\@tempswatrue
	\@citex}{\@tempswafalse\@citex[]}}
\def\citelow{\@cghifalse\@ifnextchar [{\@tempswatrue
	\@citex}{\@tempswafalse\@citex[]}}
\def\@cite#1#2{{$\null^{#1}$\if@tempswa\typeout
	{IJCGA warning: optional citation argument 
	ignored: `#2'} \fi}}

\def\@refcitex[#1]#2{\if@filesw\immediate\write\@auxout
	{\string\citation{#2}}\fi
\def\@citea{}\@refcite{\@for\@citeb:=#2\do
	{\@citea\def\@citea{, }\@ifundefined
	{b@\@citeb}{{\bf ?}\@warning
	{Citation `\@citeb' on page \thepage \space undefined}}
	\hbox{\csname b@\@citeb\endcsname}}}{#1}}

\def\@refcite#1#2{{#1\if@tempswa\typeout
        {IJCGA warning: optional citation argument
	ignored: `#2'} \fi}}

\def\refcite{\@ifnextchar[{\@tempswatrue
	\@refcitex}{\@tempswafalse\@refcitex[]}}

\def\pmb#1{\setbox0=\hbox{#1}
	\kern-.025em\copy0\kern-\wd0
	\kern.05em\copy0\kern-\wd0
	\kern-.025em\raise.0433em\box0}

\def\fnt#1#2{\footnotetext{\kern-.3em
	{$^{\mbox{\scriptsize #1}}$}{#2}}}

\font\tenrm=cmr10
\font\tenit=cmti10 
\font\tenbf=cmbx10
\font\bfit=cmbxti10 at 10pt
\font\ninerm=cmr9

\font\eightrm=cmr8

\def\qed{\hbox{${\vcenter{\vbox{			%HOLLOW SQUARE
   \hrule height 0.4pt\hbox{\vrule width 0.4pt height 6pt
   \kern5pt\vrule width 0.4pt}\hrule height 0.4pt}}}$}}

\begin{document}

%\runninghead{Rosu, Testing the water of
%$\ldots$} {Rosu, quantum heat bath
%$\ldots$}

%Comment (HCR): produce frazele de mai sus la inceputul fiecarei pagini

\normalsize\textlineskip
\thispagestyle{empty}
\setcounter{page}{1}

\copyrightheading{}			%{Vol. 0, No.0 (1992) 000--000}

\vspace*{0.88truein}

%\fpage{1} %%%%%%%%%%%%%%%%%%%%%%%%%%%%%%%%%%%%%%%%%%%%%%%%%%%%%%%%%%%
\centerline{\bf BLIND SPOT MAY REVEAL VACUUM RADIATION 
     }
%\centerline{\bf  }
\vspace*{0.035truein}
%\centerline{\bf MANUSCRIPTS USING COMPUTER SOFTWARE\footnote{For
%the title, try not to use more than 3 lines. Typeset the title
%in 10 pt Times Roman, uppercase and boldface.}}
\vspace*{0.37truein}
%\centerline{\footnotesize }
%\footnote{Typeset names in
%10 pt Times Roman, uppercase. Use the footnote to indicate the
%present or permanent address of the author.}}
%\vspace*{0.015truein}
%\centerline{\footnotesize\it Departamento de F\'{\i}sica, CINVESTAV-IPN,
%Apdo Postal 14-740, 07000 M\'exico D.F., Mexico}
%\baselineskip=10pt
%\centerline{\footnotesize\it City, State ZIP/Zone,
%Country\footnote{State completely without abbreviations, the
%affiliation and mailing address, including country. Typeset in 8
%pt Times Italic.}}
%\vspace*{10pt}
\centerline{\footnotesize  from HARET ROSU}
\vspace*{0.015truein}
\centerline{\footnotesize\it in the Instituto de F\'{\i}sica at the
Universidad de Guanajuato, Apdo Postal E-143, 37150 Le\'on, Gto, Mexico}
%\baselineskip=10pt
%\centerline{\footnotesize\it City, State ZIP/Zone, Country}
\vspace*{0.225truein}
%\publisher{(January 9, 1999)}{(later or not necessary)}

%%%%%%%%%%%%%%%%%%%%%%%%%%%
%\vspace*{0.21truein}
%\abstracts{{\bf Summary}: -
%}{}{}
%%%%%%%%%%%%%%%%%%%%%%%%%%%

%\vspace*{10pt}
%\keywords{The contents of the keywords}

\textlineskip                  %) USE THIS MEASUREMENT WHEN THERE IS
\vspace*{12pt}                 %) NO SECTION HEADING

\vspace*{1pt}\textlineskip	%) USE THIS MEASUREMENT WHEN THERE IS
%\section{General Appearance}    %) A SECTION HEADING
\vspace*{-0.5pt}
\noindent

%%%%%%%%%%%%%%%%%%%%%%%%%%%%%%%%%%%%%%%%%%%%%%
%PACS number(s):  98.80.Hw, 11.30.Pb

\noindent
%%%%%%%%%%%%%%%%%%%%%%%%%%%%%%%%%%%%%%%%%%%%%%%%%%%%%%%%%%%%%%%%%%%%%

%\newpage

%\pagebreak

%\textheight=7.8truein
%\setcounter{footnote}{0}
%\renewcommand{\thefootnote}{\alph{footnote}}

%\section{The Main Text}
\noindent
Back in the 1970s Stephen Hawking of Cambridge University in the UK made the 
theoretical discovery that small black holes are not ``completely black". 
Instead, a black hole emits radiation with a well-defined temperature that is
 proportional to the gravitational force at its surface. The finding uncovered 
a deep connection between gravity, quantum mechanics and thermodynamics. And 
later, Bill Unruh of the University of British Columbia in Canada proposed 
that quantum particles should emit thermal radiation in a similar way when 
they are accelerated. According to Unruh, a particle undergoing a constant 
acceleration would be embedded in a ``heat bath" at temperature
%%%%%%%%%%%%
$
T =\frac{\hbar}{2\pi c k}\cdot a~,
%\eqno(1.1)
$
%%%%%%%%%%%%
where $\hbar$ is the Planck constant divided by 2$\pi$, $a$ is 
the acceleration, $c$ is the speed of light and $k$ is the Boltzmann constant. 
But is it really possible to detect such radiation?
Recently, Pisin Chen of the Stanford Linear Accelerator Center and Toshi Tajima 
of the University of Texas at Austin in the US have suggested that it should 
be possible to detect the Unruh radiation emitted by electrons that are 
accelerated by high intensity lasers (1999 Phys. Rev. Lett. 83 256). The 
difficulty with detecting Unruh radiation is that enormous accelerations are 
required to produce a measurable effect. For instance, we would have to 
accelerate a particle to over $10^{20}$~m/s$^2$ to generate a temperature of
 1~K. 
Recent advances in laser research mean that lasers can now deliver subpicosecond
 pulses with petawatts of power. These could produce accelerations that are 
10$^{25}$ times greater than the acceleration due to gravity at the 
Earth's surface, 
and two orders of magnitude larger than previous experimental proposals.
At the quantum level, the vacuum is full of particles and antiparticles that 
constantly appear and disappear. The Heisenberg uncertainty principle allows 
these ``virtual" particles to exist for a very brief moment of time before they 
recombine and disappear into the vacuum again. According to Hawking if a 
particle and antiparticle are created close to the surface of a black hole, 
the strong gravitational force will pull one of the particles into the hole 
while the other escapes. Thus the black hole can produce ``radiation from 
nothing". Similarly, Unruh radiation comes from the quantum vacuum.
The curious feature about Hawking radiation is that the temperature is 
inversely proportional to the mass of the emitting source. The only black holes 
that may be detectable are ``miniholes" that may have been formed shortly after
 the Big Bang. Such black holes would have a mass of 10$^{15}$~grammes and 
would be 
smaller than a single atom.
The Unruh effect is considered slightly less esoteric, and in the 1980s several 
groups proposed experiments to detect the radiation. Unruh himself suggested 
that sound waves would propagate in a supersonic fluid flow in the same way 
that quantum fields propagate in the vicinity of a black hole. And shortly 
afterwards the late John Bell and Jon Leinaas of the University of Oslo in 
Norway suggested that the Unruh effect would alter the motion of particles at 
high-energy accelerators. A more realistic experiment was suggested by Joseph 
Rogers, now at Cornell University in the US, in which an electron confined 
by electric and magnetic fields in a so-called Penning trap would give a signal.
Meanwhile Eli Yablonovitch, now at the University of California at Los Angeles 
proposed that Unruh radiation would be produced when a gas is suddenly ionized 
and turns into a plasma. And Simon Darbinyan of the Yerevan Physics Institute 
in Armenia and co-workers suggested that Unruh radiation could be produced by 
a beam of particles propagating through channels in a crystal lattice.
In all of these experiments, however, the Unruh signal would be buried beneath 
a much larger background signal, a problem that Chen and Tajima have managed 
to circumvent. Moreover, in their scheme an electron can be instantly 
accelerated and decelerated in every laser cycle.
Chen and Tajima present simple calculations for the acceleration produced by a 
standing wave produced by two counter-propagating, ultra-intense laser pulses. 
They propose to detect the Unruh radiation from a minute change to the classical
 Larmor radiation emitted when an electron is accelerated.
Despite the high acceleration produced in a petawatt laser, the power of the 
emitted Unruh radiation is several orders of magnitude smaller than the power 
of the Larmor radiation. However, Chen and Tajima calculated the angular 
distribution of both types of radiation and found a ``blind spot" where the
 Unruh signal would dominate over the Larmor radiation (see figure).
Although appealing, the proposal of Chen and Tajima is based on several 
assumptions that may not actually be true. For example, they assume that the 
electron has a well-defined acceleration, velocity and trajectory. Moreover, 
in 1988 Alexander Nikishov and Vladimir Ritus of the Lebedev Physical 
Institute in Moscow suggested that the Unruh heat-bath concept could not be 
tested using charged particles in an electric field. They argued that the 
particle and antiparticle pairs created from the vacuum would encounter a
 varying acceleration field over a short timescale, whereas Unruh radiation is 
related to constant accelerations only. And at a recent workshop on the
 quantum aspects of beam physics, John Jackson from the University of 
California at Berkeley warned against trying to interpret conventional
 phenomena in terms of Unruh radiation. Nevertheless it is challenging to 
look for new ways to test quantum field theory that may give an insight into 
the physical origin of Hawking radiation.

\bigskip

%%%%%%%%%%%%%%
\vskip 1ex
\centerline{
\epsfxsize=260pt
\epsfbox{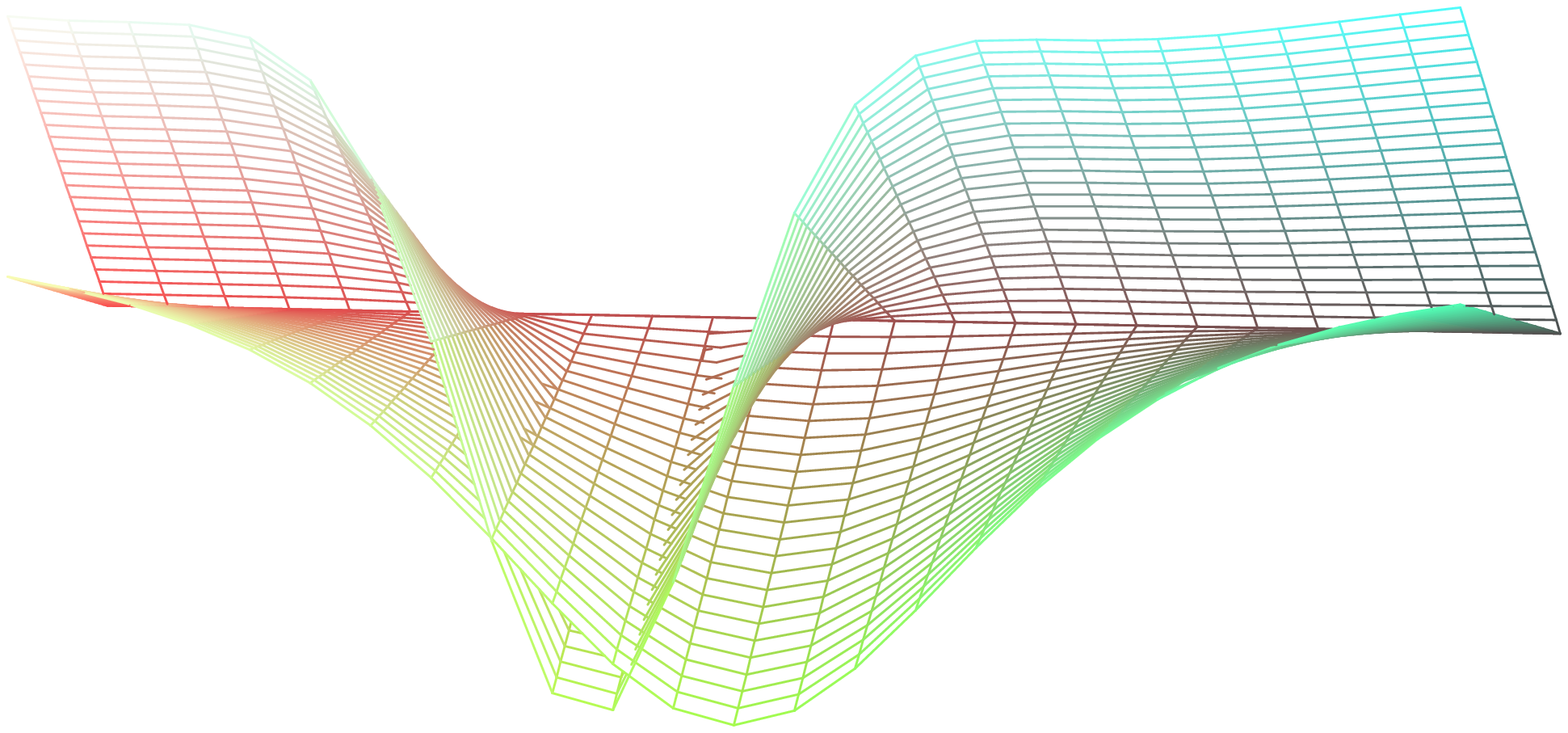}}
\vskip 2ex
%%%%%%%%%%%%%%%%%%

\bigskip

Figure: The angular distributions of the Larmor (top) and Unruh radiation 
(bottom) emitted by an accelerated electron. In general the power of the 
background Larmor radiation is much greater than the Unruh signal, but there 
is a small ``blind spot" where the Unruh radiation dominates.

\end{document}